\documentclass[unsortedaddress,aps,pre,preprint,show pacs]{revtex4}
\usepackage{graphicx}
\usepackage{dcolumn}
\usepackage{amsmath}
\usepackage{amssymb}
\usepackage{amsfonts}
\usepackage{bm}
\usepackage{color}
\newcommand{\be}{\begin{equation}}
\newcommand{\ee}{\end{equation}}
\newcommand{\ba}{\begin{eqnarray}}
\newcommand{\ea}{\end{eqnarray}}

\begin{document}
\title{Self-assembly in a model colloidal mixture of dimers\\and spherical particles}
\author{Santi Prestipino\footnote{Corresponding author. Email: {\tt sprestipino@unime.it}}, Gianmarco Muna\`o%
, Dino Costa%
, and Carlo Caccamo%
}
\affiliation{Dipartimento di Scienze Matematiche ed Informatiche, Scienze Fisiche e Scienze della Terra, Universit\`a degli Studi di Messina, Viale F. Stagno d'Alcontres 31, 98166 Messina, Italy}
\date{\today}
\begin{abstract}
We investigate the structure of a dilute mixture of amphiphilic dimers and spherical particles, a model relevant to the problem of encapsulating globular ``guest'' molecules in a dispersion. Dimers and spheres are taken to be hard particles, with an additional attraction between spheres and the smaller monomers in a dimer. Using Monte Carlo simulation, we document the low-temperature formation of aggregates of guests (clusters) held together by dimers, whose typical size and shape depend on the guest concentration $\chi$. For low $\chi$ (less than $10\%$), most guests are isolated and coated with a layer of dimers. As $\chi$ progressively increases, clusters grow in size becoming more and more elongated and polydisperse; after reaching a shallow maximum for $\chi\approx 50\%$, the size of clusters again reduces upon increasing $\chi$ further. In one case only ($\chi=50\%$ and moderately low temperature) the mixture relaxed to a fluid of lamellae, suggesting that in this case clusters are metastable with respect to crystal-vapor separation. On heating, clusters shrink until eventually the system becomes homogeneous on all scales. On the other hand, as the mixture is made denser and denser at low temperature, clusters get increasingly larger until a percolating network is formed.
\end{abstract}
\maketitle

\section{Introduction}

Supra-molecular self-assembly, {\it i.e.}, the spontaneous formation of organized structures from pre-existing simpler building blocks, is among the prominent features of soft matter~\cite{Likos}. At present, the list of morphologies observed in experiments includes micelles of various shapes, lamellae, gyroids, networks, and vesicles~\cite{Israelachvili,Thorkelsson}. The ability to control self-assembly is an important aspect in the fabrication of synthetic materials tailored for targeted applications~\cite{Yethiraj1,Wang}.

In simulations, self-assembly has often been studied in systems of ``patchy'' particles~\cite{Zhang,Glotzer,Bianchi1,Doppelbauer,Matthews,Fusco:13,Munao:13,Frank:13,Munao-JCP:14,Kegel:15}, {\it i.e.}, spherical particles with a handful of interaction spots on their surface, offering the advantage of a flexible angle-dependent interparticle potential that can be tuned to attain the desired equilibrium structures. In the last decade, these systems have gained utmost importance owing to the possibility to synthesize colloidal particles mimicking a wide assortment of patchy particles~\cite{Kraft:11,Wang,Kraft:16}. Interestingly, a wide variety of patterns is also observed in simpler systems of Janus dimers~\cite{Janusdumbbell1,Janusdumbbell2,Sun:14,Munao:14,Avvisati-JCP,Avvisati:15,Avvisati-Soft,Barbosa:15,Hatch1,Hatch2}, {\it i.e.}, heterodimers where one monomer is lyophilic and the other one is lyophobic. In essence, Janus dimers represent the ``molecular'' analogue of Janus spheres~\cite{janusprl,Larson:14,Sun:16,Figueroa:16}; like their ``atomic'' counterparts, Janus dimers are able to organize into various super-structures, such as micelles and bilayers, as well as to separate into vapor and liquid~\cite{Munao:JPCM}. 

Very recently, we have proposed heteronuclear dimers as encapsulating agents for spherical particles~\cite{Munao:capsule}. Aside from its fundamental interest, the reversible formation of capsules~\cite{Li,Malik} around a dissolved target species (``guest'') is an issue of growing importance in the pharmaceutical field, where it finds application in drug delivery~\cite{Kataoka,Singh,Kumari}, and for food industry as well, as a tool to improve the delivery of bioactive substances into foods~\cite{McClements,Dordevic}. In our model~\cite{Munao:capsule}, a dimer consists of a pair of tangent hard spheres of different size, with an additional square-well attraction between small monomers and spheres. With a minimum number of parameters this model allows for the formation of non-trivial structures at equilibrium, which depend on the relative size of dimers and spheres. For a mixture of a small fixed density, we have demonstrated~\cite{Munao:capsule} the formation of coating layers of dimers onto guests (``capsules''), at least provided that the guest concentration is sufficiently low. This basic model is meant to represent, within an implicit-solvent description, a colloidal dispersion made of a guest species and a Janus dimer where the small monomer is at once lyophobic and shows a strong affinity for the guests. Such a system may be employed, for instance, to describe at a basic level the coating of dispersed proteins by colloidal particles, where the proteins are represented in terms of either central potentials or site-distributed interactions~\cite{Abramo1,Abramo2,Pellicane}. More important, colloids can be designed with the characteristics of our model~\cite{Wang,Kraft:11,Skelhon,Granick:15,Kraft:16}, implying that its self-assembly can be probed experimentally, at least in principle.

In this paper we explore by Monte Carlo simulation the self-assembly of the same mixture studied in Ref.\,\cite{Munao:capsule}, allowing the overall density and the sphere concentration to span over a wider range than previously considered. In the present analysis, the sphere size is fixed to that of the large monomers. By this choice, we aim to represent a generic colloid where the dimers are tailored to roughly match the size of the guest to be captured. We are interested in working out the full self-assembling behavior of the system as a function of the sphere concentration $\chi$ at low density. In essence, our simulations show the existence of aggregates of spheres held together by dimers, reaching the largest size for a moderate value of $\chi$. By performing a detailed microscopic analysis, we identify and characterize the clusters present in the system, from capsules to more open structures of tubular shape. In one case only ({\it i.e.}, at equimolar concentration and a moderately low temperature), particles spontaneously self-assembled into a lamellar aggregate, which demonstrates the existence of a competition between organized structures of different nature. Finally, upon increasing the system density at fixed concentration we observe the formation of a percolating network of guests (``gel'').

Mesoscale structures are usually the outcome of competing short-range attractive and long-range repulsive interactions~\cite{Zhuang}; they are observed in such diverse fields as magnetic alloys~\cite{Seul}, Langmuir films~\cite{Keller}, and protein solutions~\cite{Stradner}. Characterization of clusters is also relevant to nucleation (see, e.g., Refs.\,\cite{Prestipino2,Prestipino3} and references therein), even though in this case they are ephemeral aggregates rather than permanent structures. Finally, spherical, cylindrical, tetragonal, and other more exotic (but metastable) clusters~\cite{Prestipino4} also occur in simple fluids, although exclusively within a two-phase coexistence region. In general, the formation of mesoscopic aggregates in fluids with competing interactions supersedes liquid-vapor coexistence, leading instead to microphase separation (also dubbed intermediate-range order; see, for example, Refs.\,\cite{Liu1,Ciach,Almarza,Zhuang}). We have computed an effective guest-guest potential, whose profile changes with thermodynamic parameters in line with the aggregates found in simulation: consistently, at low temperature this potential shows short-range attractive and long-range repulsive components. The repulsive part provides the limiting factor for the growth of clusters; in the present case, it results from the segregation of attractive sites in the cluster interior.

The outline of the paper is the following. After describing the model and the method in Section 2, we present and discuss our results in Section 3. Some concluding remarks and perspectives are finally given in Section 4.

\section{System and method}
\setcounter{equation}{0}
\renewcommand{\theequation}{2.\arabic{equation}}

\subsection{Model}

In our system, a dimer is modeled as a pair of tangent hard spheres with different diameters, $\sigma_1$ and $\sigma_2$. We hereafter assume a fixed size ratio $\sigma_2/\sigma_1$ of 3, which ensures that the dimers effectively isolate the guest from the solution in those cases where capsules form~\cite{Munao:capsule}. Guest particles (species 3) are represented as hard spheres of diameter $\sigma_3=\sigma_2$. All particle interactions are hard-sphere-like with additive diameters $\sigma_{\alpha\beta}=(\sigma_\alpha+\sigma_\beta)/2$, except for the interaction between the smaller monomer in a dimer (species 1) and a guest sphere, which is given by a spherically-symmetric square-well potential:
\ba
u_{13}(r)=\left\{
\begin{array}{ll}
\infty & \quad{\rm for}\,\,\,r<\sigma_{13} \\
-\varepsilon & \quad{\rm for}\,\,\,\sigma_{13}\le r\le\sigma_{13}+\Delta\\
0 & \quad{\rm otherwise,}
\end{array} \right.
\label{eq2-1}
\ea
$\Delta=\sigma_1$ being the square-well width. In the following, $\sigma_2$ and $\varepsilon$ are taken as units of length and energy respectively, which in turn defines a reduced distance $r^*=r/\sigma_2$ and a reduced temperature $T^*=k_{\rm B}T/\varepsilon$ ($k_{\rm B}$ being Boltzmann's constant). Finally, we denote by $N_1$ and $N_3$ the number of dimers and guests, respectively. Hence $N=N_1+N_3$ is the total number of particles.

\subsection{Simulation}

Monte Carlo (MC) simulations have been performed in the canonical ensemble, using the standard Metropolis algorithm. Canonical conditions seem more appropriate in our context, since they allow to represent a mixture of dimers and spheres in the realistic setting where the relative amount of the two species is fixed from the outset. In our runs, the number of particles of each species is chosen according to the prescribed concentration of guests, $\chi=N_3/N$ with $N_3=400$; only for the lowest analyzed concentration ($\chi=10\%$) we have set $N_3=200$. A small number of runs for larger samples have also been carried out, in order to check the irrelevance of finite-size effects. Particles are initially distributed at random in a cubic simulation box of volume $V$, with periodic conditions at the box boundary; MC evolution from such a configuration thus mimics relaxation of the system to equilibrium after a quench from very high temperature (we checked in a few cases that gradual cooling of the system from high temperature gives essentially the same results). For most of our runs we work with an overall system density $\rho^*=N\sigma_2^3/V=0.05$, a relatively low value. Moreover, we have considered one case ($\chi=33\%$) where the system is progressively made denser at fixed temperature.

A single MC cycle consists of $N$ trial single-particle moves. Depending on the type of particle, one trial move is a simple translation or a random choice between center-of-mass translation and rotation about a coordinate axis. The acceptance rule as well as the schedule of the moves are designed in such a way that detailed balance is satisfied. The maximum random shift and rotation are adjusted during equilibration so as to keep the ratio of accepted to total number of moves close to $50\%$ for $T^*\le 0.20$, and to $60\div 70\%$ otherwise. To speed up execution we have implemented linked lists in our computer code, which turns out to be especially useful at low temperature ($T^*\le 0.15$) where equilibration runs are longer. For $\chi=33\%$ we have also evolved the system with the aggregation-volume-bias (AVB) MC algorithm~\cite{Chen}. While the AVB simulation converged quickly than the standard MC simulation for $T^*\ge 0.20$, the AVB algorithm turns out to be far less efficient than the Metropolis algorithm for $T^*=0.10$. Rejection-free moves~\cite{Liu2} is another alternative to bare Metropolis, which could be considered for future studies.

In order to decide whether the system has reached equilibrium for assigned values of $\chi$ and $T$ we look at the evolution of the potential energy $U$ with the number of MC cycles performed: a total energy fluctuating around a fixed value for long is the hallmark of stable equilibrium. For $T^*=0.20$ or larger, $5\times 10^7$ MC cycles are well enough for attaining equilibrium, while we had to generate from 3 to $9\times 10^8$ cycles (depending on $\chi$) before computing the structure quantities for $T^*=0.10$. Indeed, only for this temperature (which is the lowest temperature investigated) the value of $U$ needed a conspicuous number of cycles to level off.

In the production runs, which are typically $10^8$ cycles long, we compute the radial distribution functions (RDF) $g_{\alpha\beta}(r)$ (with $\alpha,\beta=1,2,3$ denoting particle species), updating the relative histograms every 100 cycles. We also carry out a cluster analysis by first identifying at regular times ({\it i.e.}, every 1000 cycles) and classifying as a function of size connected structures of guest spheres by the Hoshen-Kopelman algorithm~\cite{Hoshen} adapted to continuous space. Any such structure clearly represents the backbone of a cluster. The criterion used to qualify two spheres that are close to each other in space as bound is a relative distance smaller than the abscissa $r_{\rm min}$ of the $g_{33}$ first minimum for $T^*=0.10$ ($r^*_{\rm min}$ varies in the interval $1.21$-$1.35$ as $\chi$ ranges from $10\%$ to $80\%$). Then, for each cluster we compute its ``size'', namely the number of spheres it contains, the binding energy $E_b$ (that is, the number of 1-3 contacts), and also count the number $N_{\rm dim}$ of dimers hosted in the cluster ({\it i.e.}, those dimers that are bound with at least one sphere of the cluster). The analysis is completed by determining the statistical distribution of the number $n_{\rm NN}$ of guests that are nearest neighbors to the same sphere and the distribution of the angle $\alpha$ formed by two 3-3 bonds having one sphere in common. 

\section{Results}
\setcounter{equation}{0}
\renewcommand{\theequation}{3.\arabic{equation}}

\subsection{Energy per guest particle}

In our previous study~\cite{Munao:capsule}, we found that guests with size comparable to that of dimers ($\sigma_3=\sigma_2/2$ and $\sigma_3=\sigma_2$, with $\sigma_1=\sigma_2/3$) can be encapsulated for all concentrations up to 20\%, at least provided that the temperature is sufficiently low ($T^*=0.15$ or smaller, see Table 3 of Ref.\,\cite{Munao:capsule}). As the guest size increases, progressively lower concentrations are required to obtain encapsulation. For large guest sizes and not too low concentrations ($\sigma_3=3\sigma_2$ and $\chi\geq 10\%$), other interesting self-assembly behaviors were observed, such as the neat separation between a guest-rich and a guest-poor phase. However, the exploration of the thermodynamic space in Ref.\,\cite{Munao:capsule} was rather limited, and no specific analysis was attempted to analyze in more quantitative terms the mesoscopic structures observed at equilibrium. In the following, we fill this gap for a mixture of dimers and spheres of comparable sizes ($\sigma_3=\sigma_2$).

%
%
\begin{figure*}
\begin{center}
\begin{tabular}{cc}
\includegraphics[width=8.0cm]{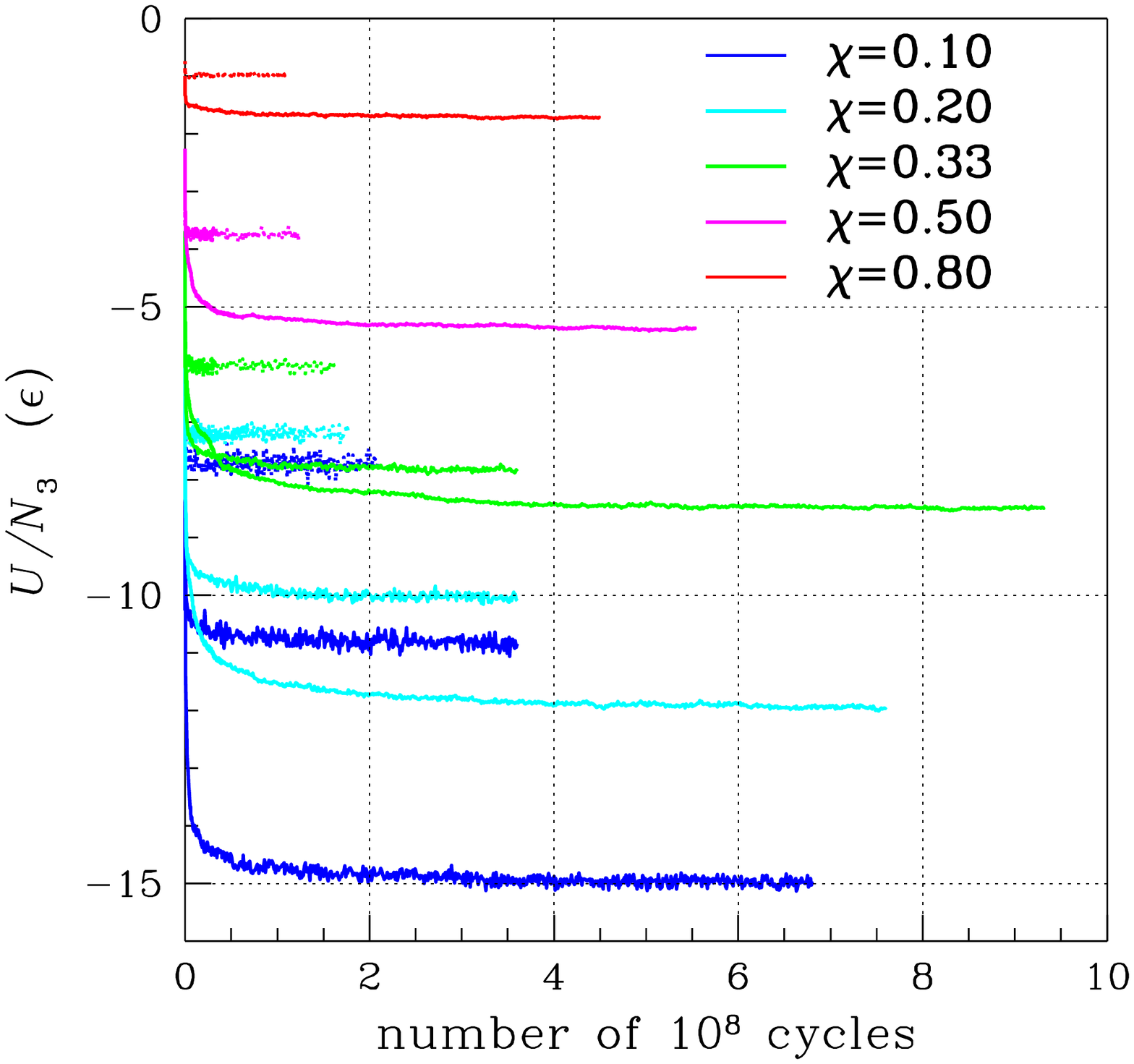} &
\includegraphics[width=8.0cm]{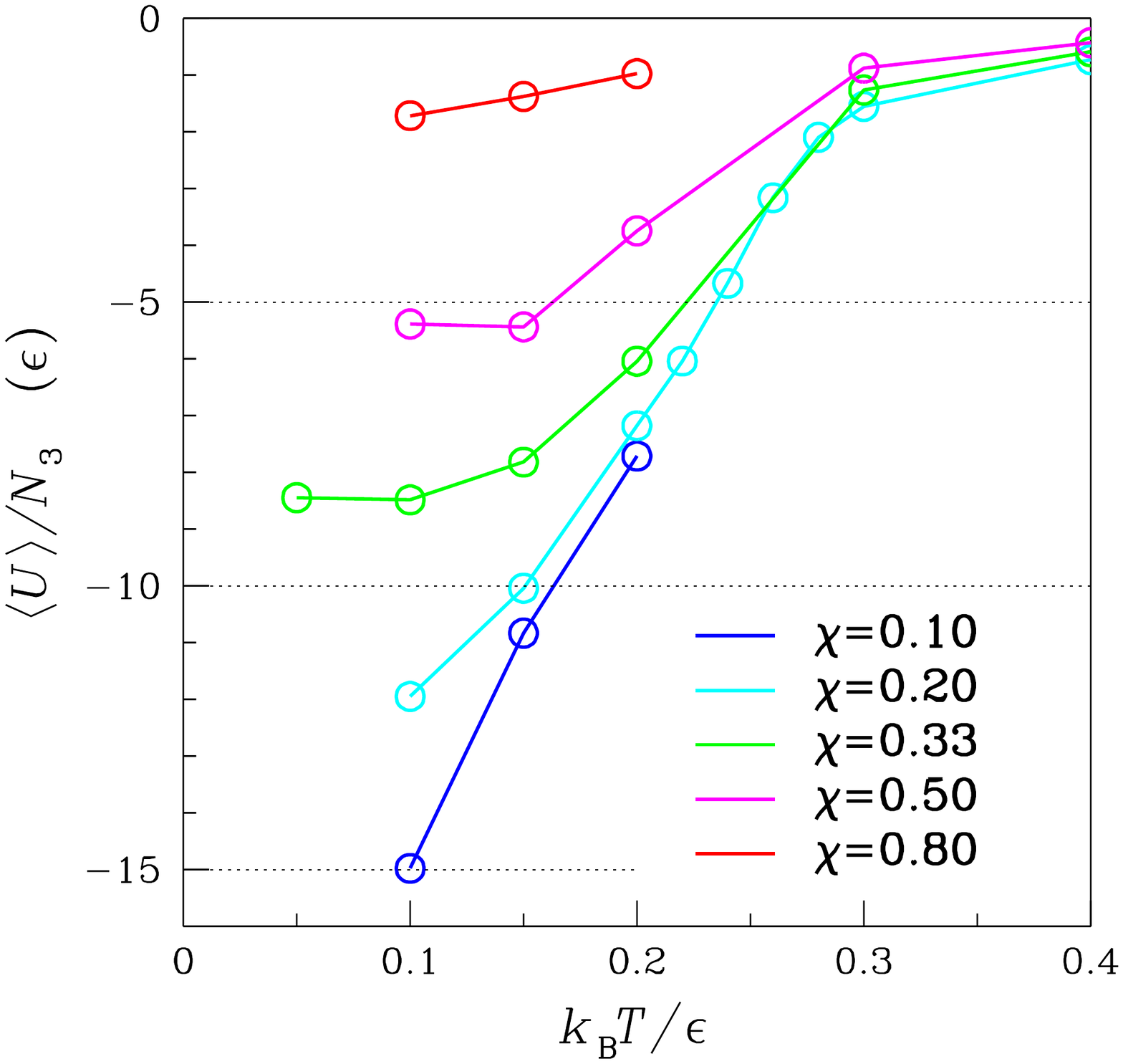}
\end{tabular}
\caption{Left: MC evolution of the potential energy $U$ per sphere, for various $\chi$ (see the legend) and for $T^*=0.10,0.15,0.20$ from bottom to top (for $\chi=50\%$ and $80\%$, the case $T^*=0.15$ is omitted). Right: average potential energy per sphere as a function of temperature. Error bars are smaller than the symbols size.}
\label{fig1}
\end{center}
\end{figure*}

We first comment on the simulation results for a system of density $\rho^*=0.05$. In the left panel of Fig.\,1 we report some representative cases of $U$ evolution in the course of the simulation run. While equilibrium is attained relatively quickly for $T^*\ge 0.20$, thermal relaxation is slower for $T^*=0.15$, and even more so for $T^*=0.10$, where we needed to produce several hundred million cycles before $U$ could level off. For example, for $\chi=33\%$ energy roughly stabilizes only after 7-$9\times 10^8$ MC cycles, while relaxation is somewhat faster for the lowest and highest concentrations investigated in this work. After the $U$ curve has flattened to a sufficient degree we compute the average potential energy $\langle U\rangle$ over a trajectory slice made of $10^8$ cycles. This quantity is plotted in the right panel of Fig.\,1 as a function of temperature for the various concentrations. At every temperature, $\left|\langle U\rangle\right|$ is higher the lower the concentration of guests. Indeed, as it appears from the subsequent analysis, when few guests are present, dimers tend to coalesce around them, at least so long as space permits, this way maximizing the number of attractive contacts. Moreover, all $\langle U\rangle$ curves tend to merge near $T^*=0.25$, which may thus be taken as an estimate of the maximum temperature for which well-definite aggregates occur in the system. For low concentrations, the number of dimers that are bound to a single sphere grows substantially on cooling; apparently, this number has not yet saturated for $T^*=0.10$. For $\chi=50\%$, the value of $\langle U\rangle$ slightly rises as we turn from $T^*=0.15$ to 0.10, indicating that longer equilibrations would be necessary for $T^*=0.10$, and the same conclusion applies for $\chi=33\%$ with regard to the transition from 0.10 to 0.05. We further point out that the energy plots for larger samples with $N=4000,\chi=10\%$ and $N=2400,\chi=33\%$ are hardly distinguishable from the corresponding plots for $N=2000$ and $N=1200$, respectively, hence finite-size effects are negligible in the present case.

Visual inspection indicates that finite-size aggregates of guests glued together with dimers (``clusters'') are formed at low $T$ for all concentrations. These clusters have also been observed in our previous work on the same model~\cite{Munao:capsule}, but no systematic study of their structure was attempted there. The nature of such clusters is various: clusters have the form of capsules for $\chi=10\%$ or lower~\cite{Munao:capsule}, while they are more elongated and polydisperse for higher concentrations (a thorough analysis of the cluster shape and structure as a function of $\chi$ and $T$ is provided in the next Section 3.2). Clearly, the value of $\Delta$ in Eq.\,(\ref{eq2-1}) is crucial to determine what kind of aggregates are observed at low temperature (a smaller, yet non-zero $\Delta$ would result in a lower $T^*$ threshold for cluster formation and, in addition, in a slower relaxation to equilibrium). In one thing all clusters are equal: once they have ceased to grow and their shape has become relatively stable, almost all the small monomers contained are buried under the surface.

Now turning to the case of $T^*=0.10$, it is difficult to say whether true equilibrium has been reached in our simulations. As just said, after an initial stage of fast growth, the size and shape of clusters stop to change appreciably. By looking at the snapshot of the system at regular intervals, we see that its late MC evolution mostly consists of tiny adjustments of the clusters already present; the reason is that a strong attraction between dimers and guests is an obstacle to achieving local equilibration, because it prevents substantial particle reshuffling among the clusters. However, rather than merely signaling a slow approach to equilibrium, the persistent decrease of $\langle U\rangle$ at low $T$ might be the clue to a regime of very slow kinetic aggregation. We are perfectly aware that Markov-chain dynamics has little to do with the true, Hamiltonian dynamics of the model; we nevertheless expect that the equilibration stage in a MC simulation still keeps some of the characteristics held by the true non-equilibrium dynamics, as could be observed in a molecular-dynamics simulation. As is known, two distinct regimes of irreversible aggregation are found in colloids (see, {\it e.g.}, Ref.\,\cite{Lin}): (i) diffusion-limited aggregation, which occurs when colloidal particles attract each other strongly, so that the aggregation rate is solely limited by the time taken for particles to encounter each other by diffusion; (ii) reaction-limited aggregation, which instead occurs when there is still a substantial repulsion between the particles, so that the aggregation rate is limited by the time taken for two of them to overcome this barrier by thermal activation. These regimes respectively correspond to the cases of fast and slow aggregation. In our system, it is unlikely but definitely not impossible that two clusters whose attractive spots preferentially lie under the surface can meet in the course of the run and join together to form a bigger aggregate. According to the above classification, this can be recognized as (a rather extreme case of) reaction-limited aggregation.

\subsection{Clusters structure}

In Fig.\,2, the last system configuration generated in our MC runs is reported for $T^*=0.10$ and 0.20. Starting from the lowest $\chi$ investigated ($\chi=10\%$, case (a) in Fig.\,2), we note that nearly compact capsules occur in large number, in a medium of isolated dimers. Most guests are associated in pairs and fully covered with dimers in such a way that no large pores are visible in the coating layer. Hence, in a regime of concentrations lower than $\chi=10\%$ the guest particles would not be in direct contact with solvent particles (however, here only implicitly present). The overall looking of the mixture is similar in a larger system of $N_3=400$ guests and $N_1=4000$ dimers, as confirmed by a nearly identical $U$ evolution (not shown). Moving to $\chi=20\%$ (case (b) in Fig.\,2), we find larger clusters coagulating in the system in the long run, with an appreciable fraction of guest spheres exposed to their surface. The biggest aggregates now occurring are elongated curved objects, which no longer resemble spheroidal capsules. For $\chi=33\%$ (cases (c) and (d) in Fig.\,2) clusters are even larger than before, but still non-percolating (bottom-left panel). Upon doubling the temperature (bottom-right panel), clusters reduce in size and the ``mobility'' of particles increases substantially, as witnessed by the faster energy relaxation and the much larger $U$ fluctuations in the course of the run.

%
%
\begin{figure*}
\begin{center}
\begin{tabular}{cccc}
\scriptsize{(a) $\chi=10\%,\,T^\ast=0.10$}&
\scriptsize{(b) $\chi=20\%,\,T^\ast=0.10$}&
\scriptsize{(c) $\chi=33\%,\,T^\ast=0.10$}&
\scriptsize{(d) $\chi=33\%,\,T^\ast=0.20$}\\
\includegraphics[width=4.0cm]{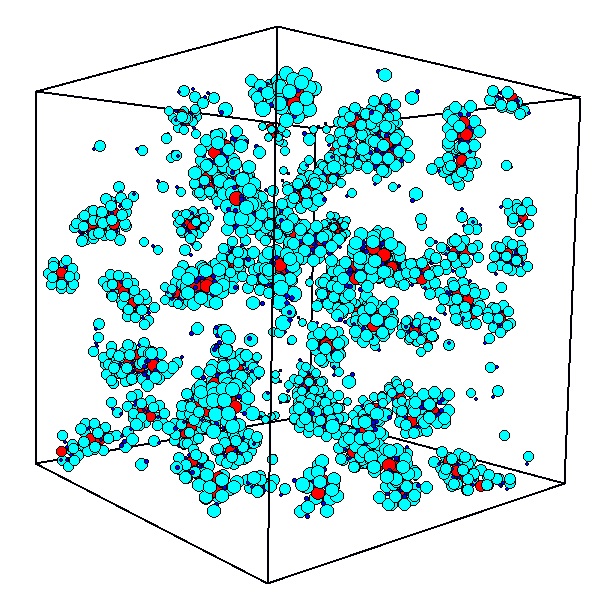}&
\includegraphics[width=4.0cm]{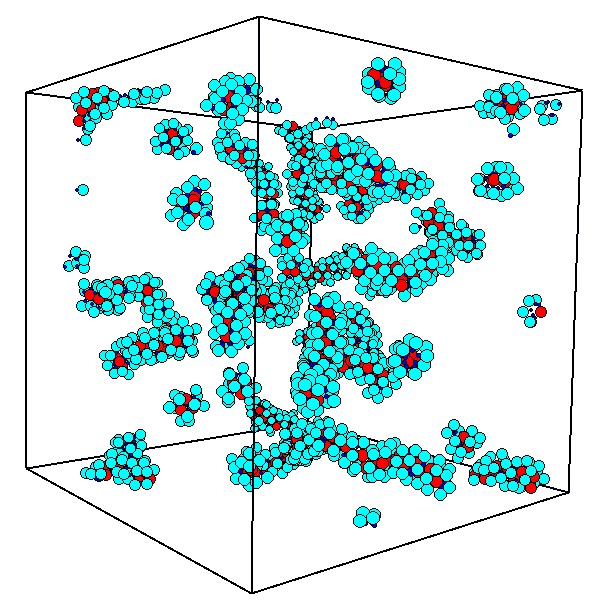}&
\includegraphics[width=4.0cm]{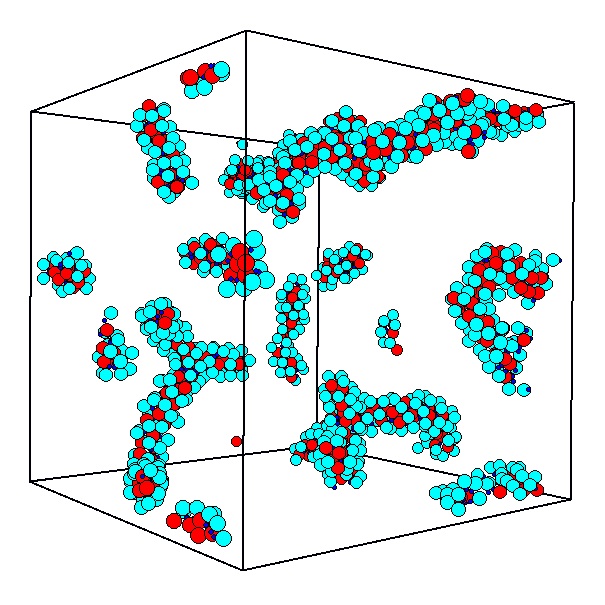}&
\includegraphics[width=4.0cm]{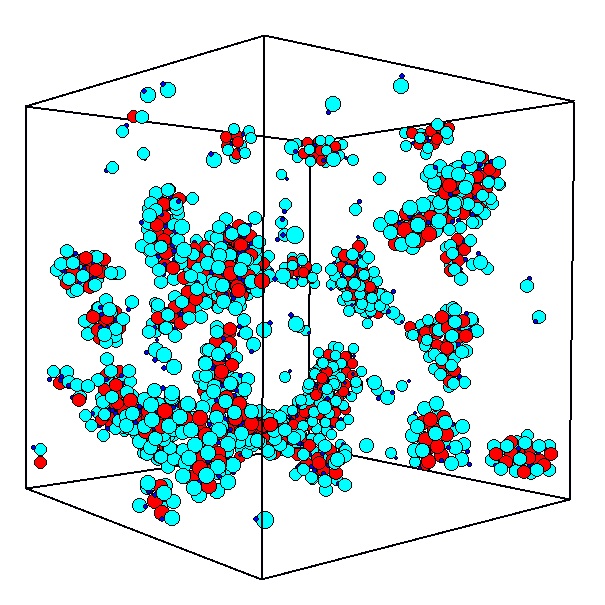}\\  \\
\scriptsize{(e) $\chi=50\%,\,T^\ast=0.10$}&
\scriptsize{(f) $\chi=50\%,\,T^\ast=0.20$}&
\scriptsize{(g) $\chi=80\%,\,T^\ast=0.10$}&
\scriptsize{(h) $\chi=80\%,\,T^\ast=0.20$} \\
\includegraphics[width=4.0cm]{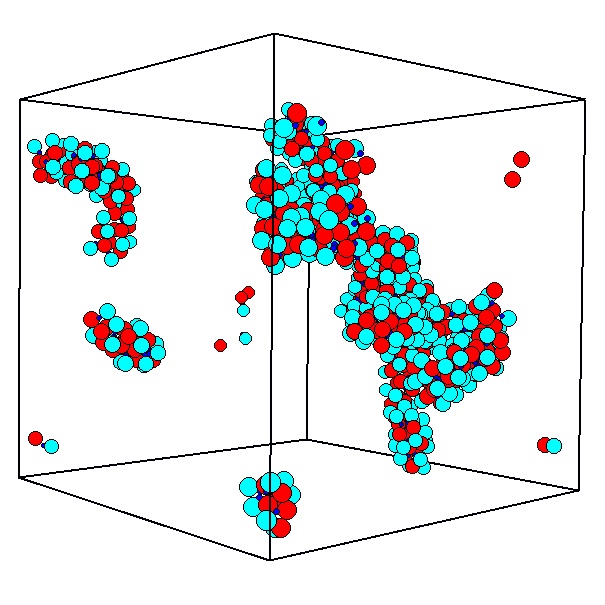}&
\includegraphics[width=4.0cm]{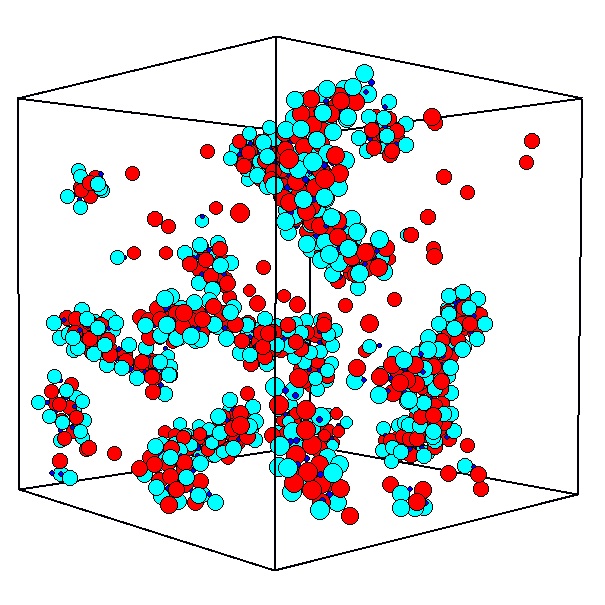}&
\includegraphics[width=4.0cm]{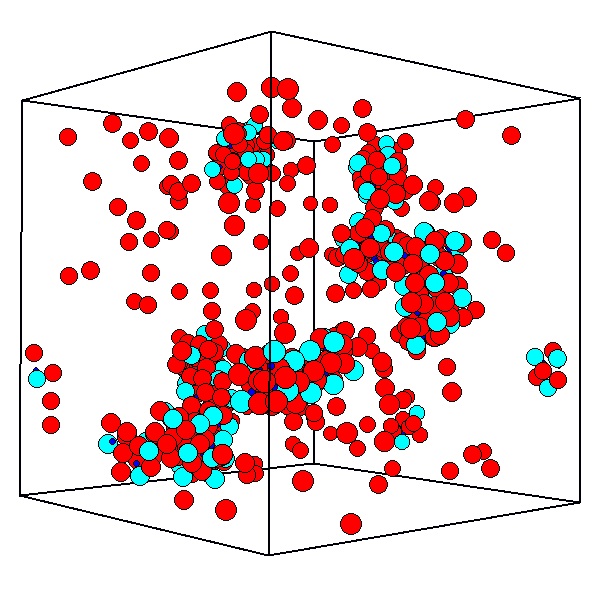}&
\includegraphics[width=4.0cm]{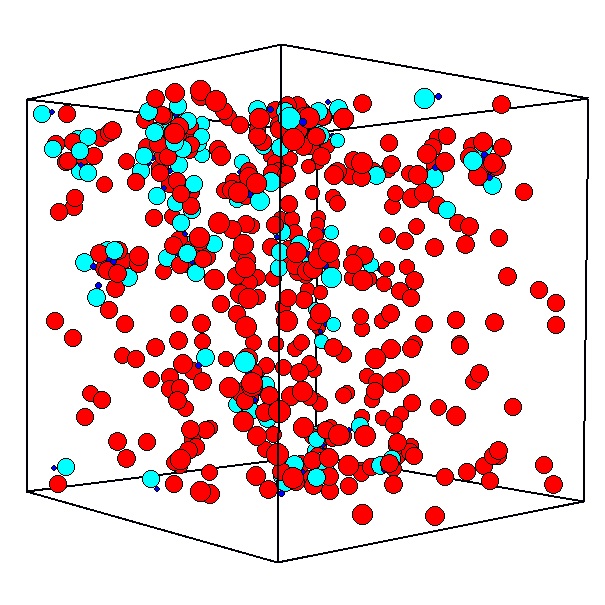}\\
\end{tabular}
\caption{Final system configurations for different $\chi$ and $T^*$ (red: spheres; cyan: large monomers; blue: small monomers).}
\label{fig2}
\end{center}
\end{figure*}

While clusters are still big at equimolar concentration (cases (e) and (f) in Fig.\,2), very much as for $\chi=33\%$, the average cluster size is definitely smaller for $\chi=80\%$ (cases (g) and (h) in Fig.\,2), where the number of unbound spheres is high even for $T^*=0.10$. The reason is clear: the few dimers present are insufficient to bind all the guests and, as a result, very large clusters are not formed. Again, heating acts against aggregation ({\it i.e.}, clusters shrink on increasing $T$), making evaporation of particles from the cluster surface easier. As a last comment, we note that ring-shaped portions occasionally appear in some cluster (see an example in Fig.\,3, which refers to $\chi=33\%$ and $T^*=0.15$). This indicates a natural tendency of clusters to bend, a property that distinguishes the present model from the fluid of one-patch particles studied in Refs.\,\cite{Munao-JCP:14,Preisler}, where tube-like aggregates are also found at low temperature.

The foregoing discussion, based on the few sketches of the system structure provided in Fig.\,2, will now be substantiated by an explicit cluster analysis, already described in Sect.\,2.2. For $T^*=0.10$ and 0.15, averages have been computed from the last $10^8$ cycles only, while for larger temperatures $5\times 10^7$ cycles are well enough for an accurate cluster statistics. We first show a comparison between simulation data relative to various concentrations at fixed temperature, $T^*=0.10$. On the left in Fig.\,4, we show the average number $N_{\rm cl}$ of clusters of equal size in a single configuration, as a function of the size $s$ (a cluster of size one is just an isolated sphere). On the right panels, we report the statistics of the number $N_{\rm dim}(s)$ of dimers hosted in a cluster, and of the number $E_{\rm b}(s)$ of 1-3 contacts.

%
%
\begin{figure}
\begin{center}
\includegraphics[width=5.0cm]{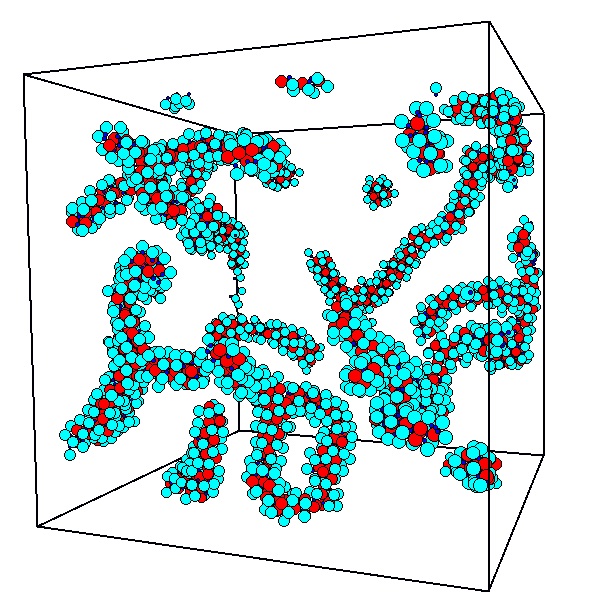}
\caption{Snapshot of a long-equilibrated system for $\chi=33\%$ and $T^*=0.15$ (here $N_1=1600$ and $N_3=800$; same notation as in Fig.\,2). A clear ring is observed in a cluster located near the bottom of the simulation box.}
\label{fig3}
\end{center}
\end{figure}

We see that statistical distributions are noisy for all values of $\chi$, which again proves that system evolution for $T^*=0.10$ is extremely slow. Gaps in the distributions reflect insufficient statistics, a problem which is more severe for the largest sizes. As far as the cluster-size distribution $N_{\rm cl}(s)$ is concerned, we find that the average number of guest spheres per cluster is about 2.5 for $\chi=10\%$, 9.3 for $\chi=20\%$, 49.6 for $\chi=33\%$, 53.6 for $\chi=50\%$, and 19.2 for $\chi=80\%$ (excluding isolated spheres from the counting). Hence we confirm that the biggest clusters occur at moderate concentration. For the same $\chi$ values the cluster-size distribution is very dispersed around the mean, while the sharp peak at 1 for $\chi=80\%$ simply reflects the existence of a large number of isolated guests at this concentration. The average number $N_{\rm dim}$ of dimers hosted in a cluster is almost linearly dependent on size, suggesting a homogeneous distribution of dimers within a cluster. The slope of $N_{\rm dim}$ vs $s$ slightly grows with decreasing $\chi$, indicating a highest density of dimers in the clusters (mostly capsules) for $\chi=10\%$. Also the cluster binding energy $E_{\rm b}$ grows linearly with size, showing a strong correlation with $N_{\rm dim}$. In Fig.\,5, the cluster statistics is reported at fixed concentration ($\chi=33\%$) as a function of temperature. Note, in particular, how the statistical accuracy of $N_{\rm cl}(s)$ substantially improves with increasing temperature, while its leading maximum moves towards lower sizes, until most guests become isolated for $T^*=0.40$.

%
%
\begin{figure*}
\begin{center}
\begin{tabular}{cc}
\includegraphics[width=8.0cm]{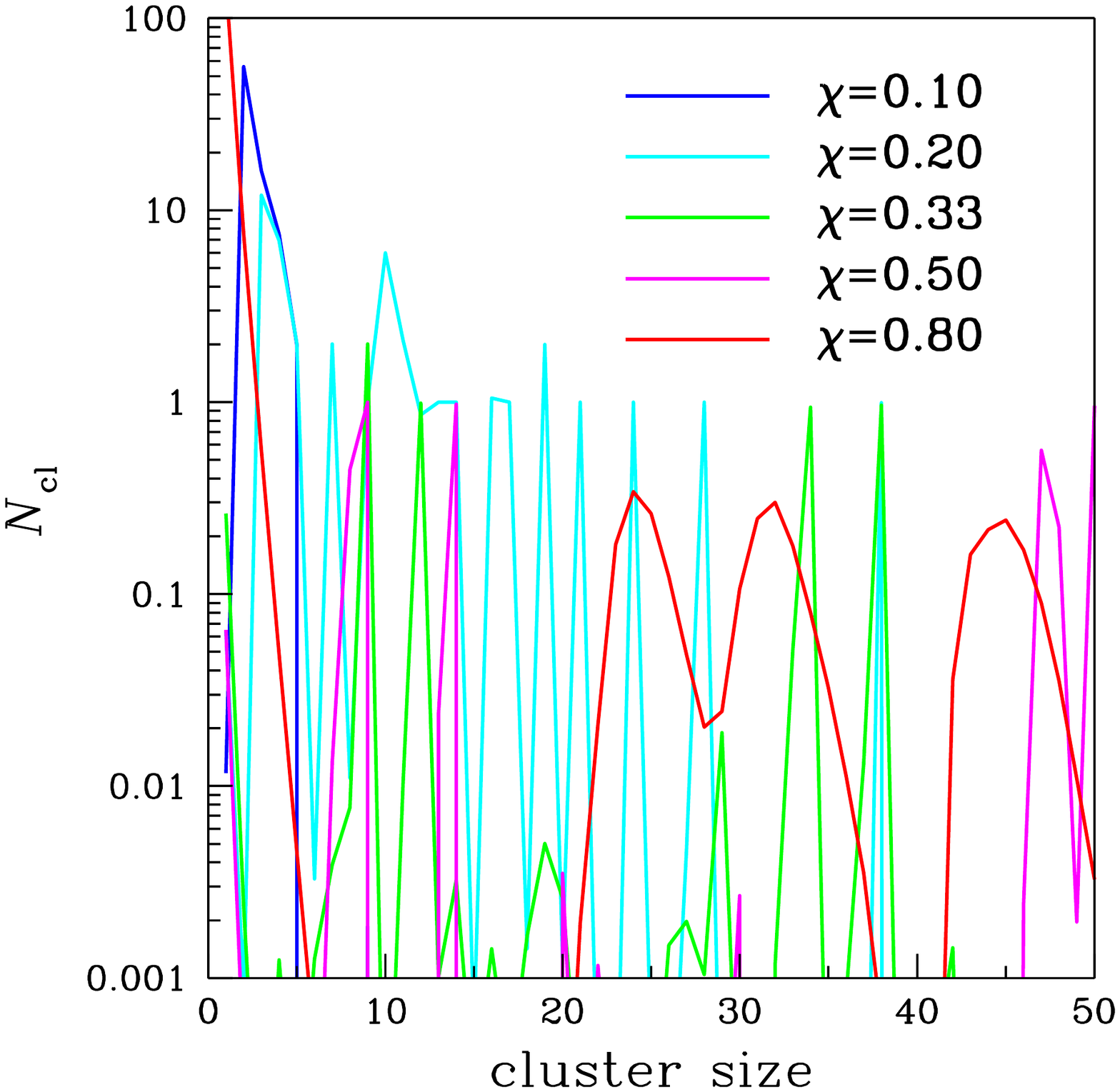} &
\includegraphics[width=8.0cm]{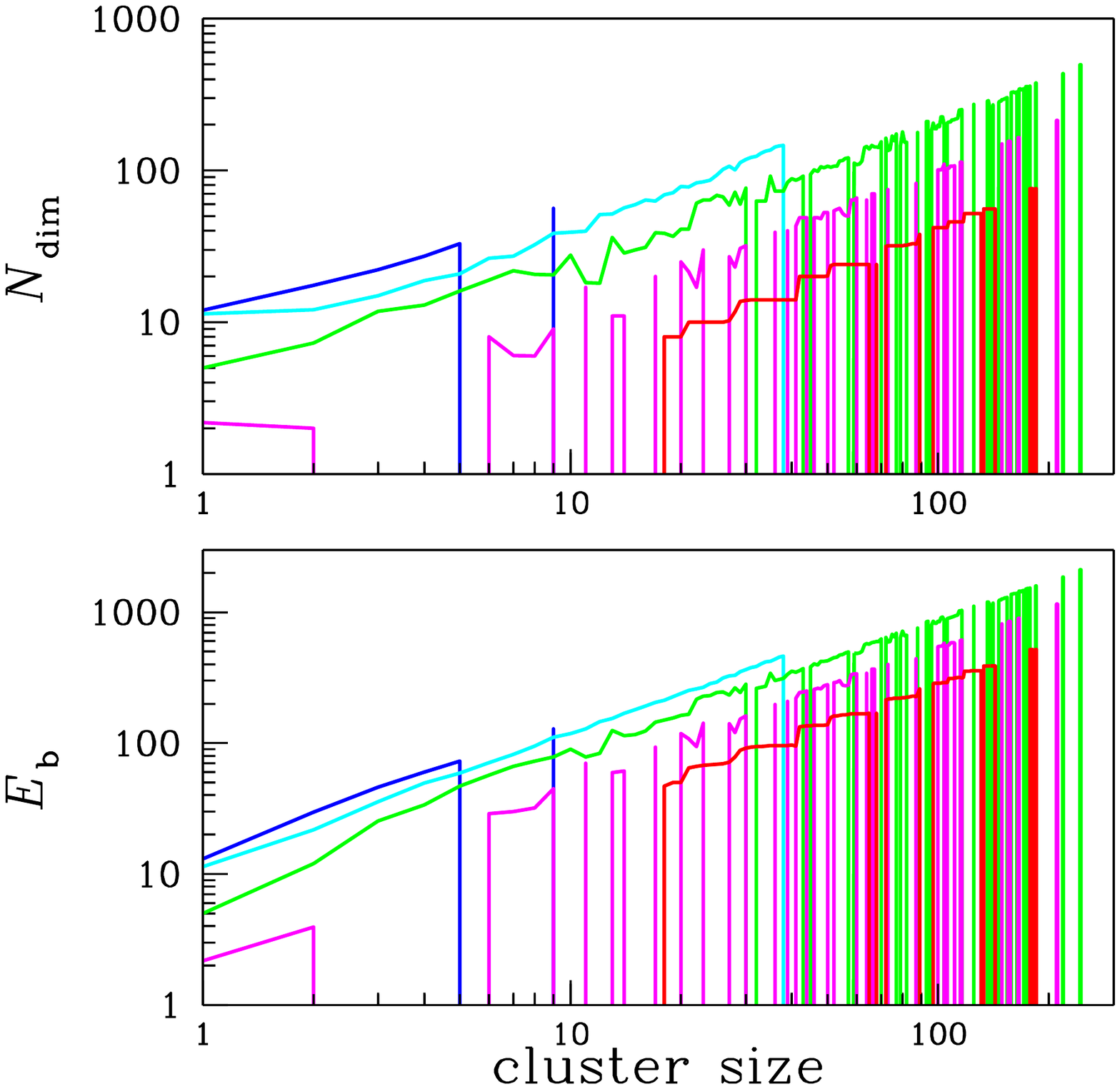}
\end{tabular}
\caption{Left: cluster-size distribution for $T^*=0.10$ and different $\chi$ (in the legend). Right: average number of dimers (top) and attractive contacts (bottom) in a cluster.}
\label{fig4}
\end{center}
\end{figure*}

%
%
\begin{figure*}
\begin{center}
\begin{tabular}{cc}
\includegraphics[width=8.0cm]{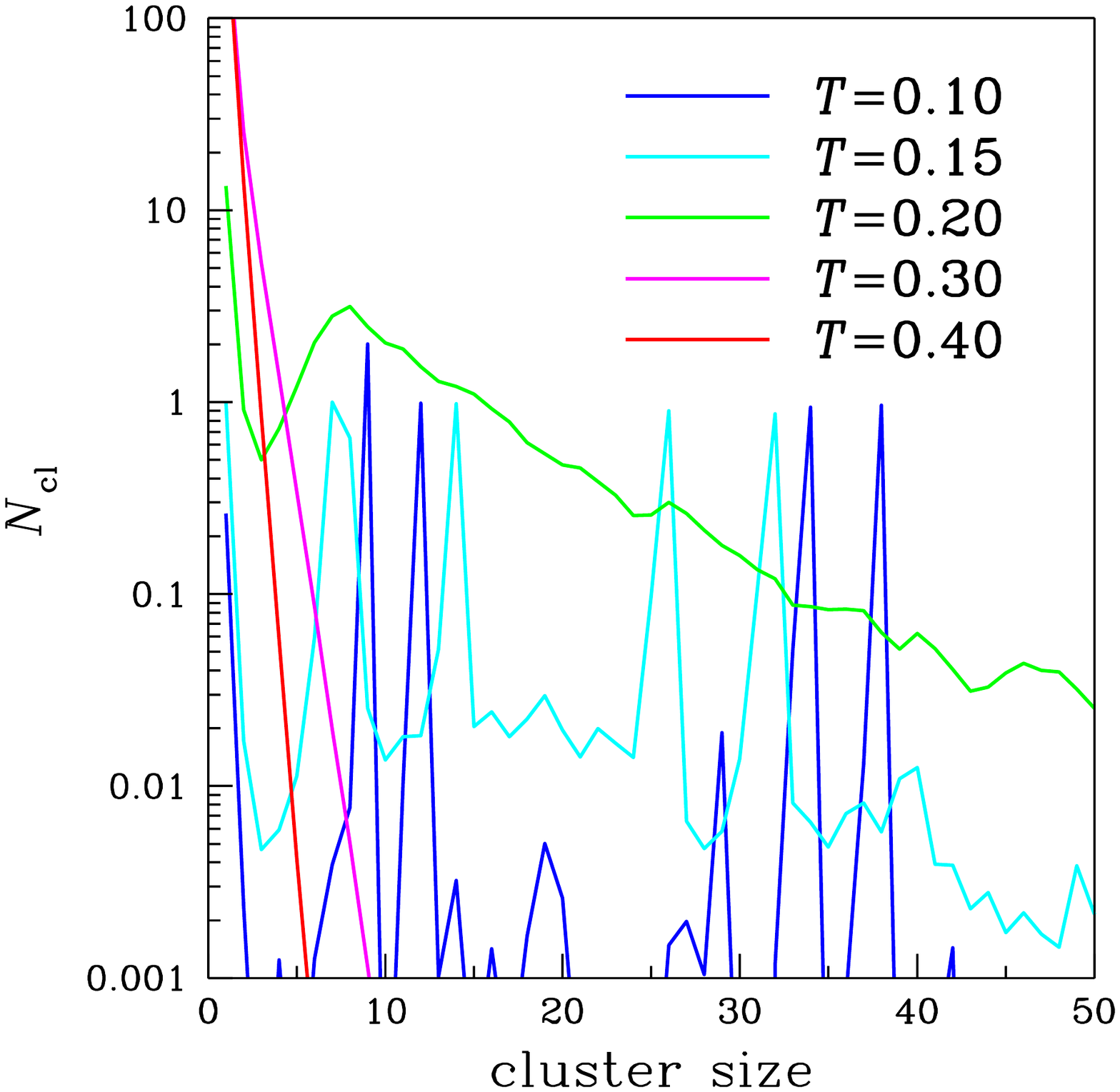} &
\includegraphics[width=8.0cm]{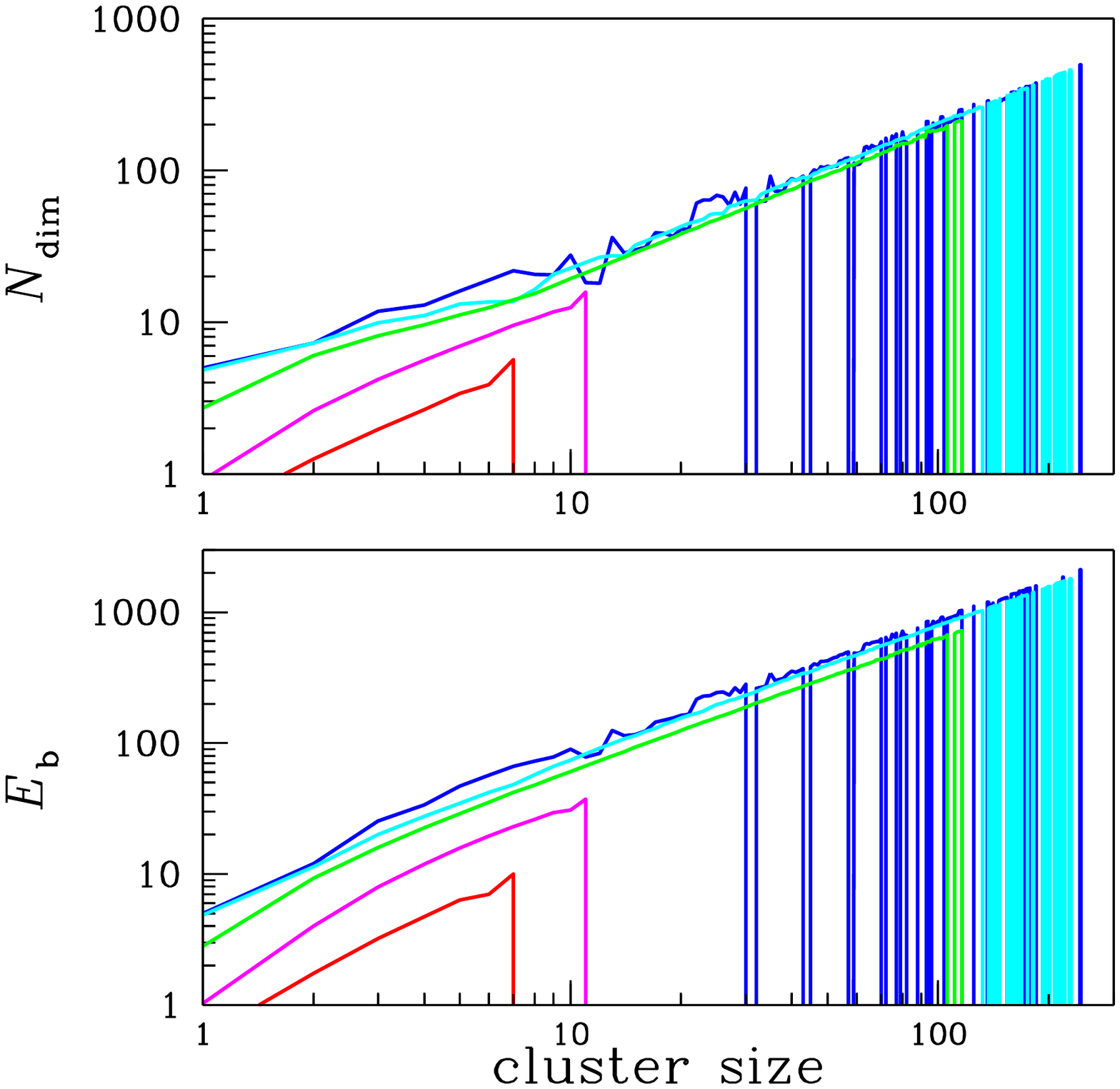}
\end{tabular}
\caption{Left: cluster-size distribution for $\chi=33\%$ and different $T^*$ (in the legend). Right: average number of dimers (top) and attractive contacts (bottom) in a cluster.}
\label{fig5}
\end{center}
\end{figure*}

As it appears from Figs.\,4 and 5, statistical uncertainties on the reported data (not shown) are significant, especially for the lowest temperatures and the biggest sizes, where instances are few and cluster lifetime is long. Indeed, a peculiarity of low-$T$ clusters is their endurance, or resistance to add/lose particles, which makes the measured $N_{\rm cl}$ a very irregular function of size. Although the features of $N_{\rm cl}(s)$ for $T^*=0.10$ are hidden by noise, we expect to find a two-peak structure at large $\chi$ where, besides the maximum at one, another shallow maximum corresponds to the most probable $s$ value in a broad distribution of sizes. As $\chi$ is reduced, the maximum at one gets strongly depressed, and has already washed away for $\chi=50\%$ (this maximum is eventually recovered upon heating, see Fig.\,5 left panel). The other maximum is roughly located near the average cluster size; hence, its abscissa would show a non-monotonous $\chi$ dependence: it first grows on approaching $\chi=50\%$ from above and then drops, until finally reaching the value of two for $\chi=10\%$. On the contrary, the average values of $N_{\rm dim}(s)$ and $E_{\rm b}(s)$ are much less uncertain (despite the many gaps present), since the relative errors are not so large as to obscure their linear behavior with $s$.

The spatial distribution of guests within the clusters can be investigated by collecting in the course of the run the statistics of a few indicators sensitive to the local sphere environment. We first present results relative to (i) the distribution $P(\alpha)$ of the angle $\alpha$ formed by two nearest guest-guest bonds, {\it i.e.}, two bonds sharing one sphere (at the angle vertex), and (ii) the statistics of the number of guest spheres that are nearest neighbors to a given sphere ({\it i.e.}, the ``coordination number'' of a sphere; clearly, all the neighbor spheres belong to the same cluster of the central sphere).

%
%
\begin{figure}
\begin{center}
\includegraphics[width=8.0cm]{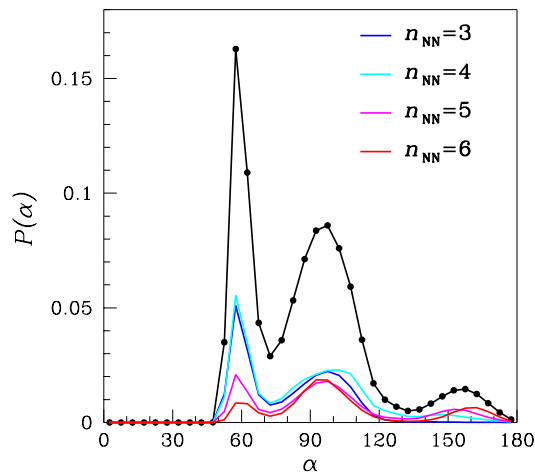}
\caption{Distribution $P(\alpha)$ for $\chi=33\%$ and $T^*=0.15$ (line with dots). We also report the separate contributions from vertex spheres characterized by a different coordination number $n_{\rm NN}$ (in the legend).}
\label{fig6}
\end{center}
\end{figure}

%
%
\begin{figure}
\begin{center}
\includegraphics[width=8.0cm]{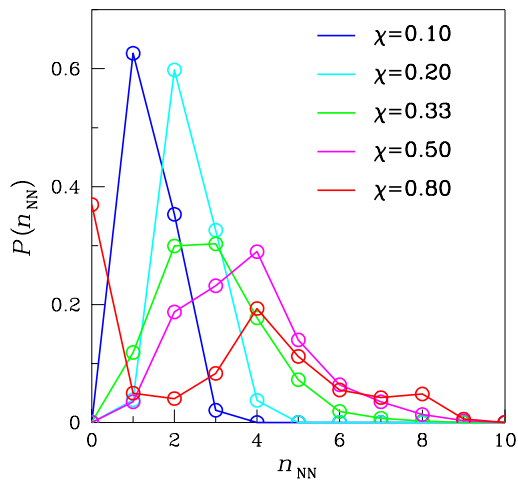}
\caption{Distribution $P(n_{\rm NN})$ of the sphere coordination number $n_{\rm NN}$, for $T^*=0.10$ and various guest concentrations $\chi$ (in the legend).}
\label{fig7}
\end{center}
\end{figure}

The $\alpha$ distribution is reported in Fig.\,6 for $\chi=33\%$ and $T^*=0.15$ (its overall shape is similar for other concentrations, being poorly defined only for small $\chi$). In the same figure, $P(\alpha)$ has been resolved into separate contributions, according to the number of neighbors owned by the vertex sphere. We see three maxima occurring in $P(\alpha)$, at $60^\circ,\approx 100^\circ$, and $\approx 160^\circ$, respectively. The maximum at $60^\circ$ is very sharp: it corresponds to three guest spheres that are in reciprocal contact. The other two maxima are much broader. The central peak is actually originated from the merging of two distinct peaks: one centered near $90^\circ$, which marks a distinct preference of the highest-coordinated spheres for a local square ordering (see more in Section 3.4 below); the other peak, occurring close to $109.5^\circ$, is the signature of a diffuse tetrahedral arrangement of fourfold-coordinated spheres within the clusters, indeed seen in the snapshots. The third maximum just results from the interposition of a guest sphere, in close contact with the vertex sphere, between the other two spheres engaged in the angle. Finally, we show in Fig.\,7 the distribution function of $n_{\rm NN}$. This number counts how many guest spheres are nearest neighbors to a given reference sphere. This number typically grows with $\chi$, moving from 1 at low concentration to $4\div 5$. The sharp maximum at zero observed for $\chi=80\%$ merely results from the large number of isolated guests.

\subsection{Effective guest-guest interaction}

The propensity of guest spheres to bind together at not-too-small $\chi$ provides evidence that the dimer-mediated interaction between guests has an attractive component. An effective pair potential between guests can be constructed as follows. We first use the Ornstein-Zernike relation~\cite{Hansennew} to compute the direct correlation function $c_{33}(r)$ from the numerical profile of $g_{33}(r)$. Then, both functions are plugged in the hypernetted-chain (HNC) closure~\cite{Hansennew} to finally obtain the guest-guest potential $\phi(r)$. We expect this HNC scheme to be valid up to at least moderate concentrations. Typical results are shown in Fig.\,8, which refers to $\chi=33\%$ (the shape of $\phi(r)$ is only weakly dependent on $\chi$). We see that a highly structured $g_{33}(r)$ ($T^*=0.15$ and 0.20) goes along with a non-trivial profile of $\phi(r)$, showing a hard-core, a short-range well, and a repulsive hump for larger distances. This profile is consistent with the existence of clusters characterized by an average cross radius of $\approx 2\sigma_2$. At sufficiently high temperature, the spatial distribution of guests is nearly homogeneous and the effective interaction becomes more hard-sphere-like.

We have not tried to use closures more sophisticated than the HNC; on the other hand, the simpler low-density approximation, {\em i.e.}, $g_{33}(r)=\exp\left\{-\beta\phi(r)\right\}$ is insufficient to obtain non-trivial structure in $\phi(r)$ at low temperature, while it reproduces the HNC outcome at high temperature. In the inset of Fig.\,8 we show the computed guest-guest structure factor $S_{33}(q)$. Intermediate-range order should be evidenced in the presence of a peak at a small $q$ value, but no such peak is present in $S_{33}(q)$. The reason of this is merely numerical: a much larger system size and much longer simulations would be needed in order to see this feature eventually appear at low temperature.

%
%
\begin{figure*}
\begin{center}
\begin{tabular}{cc}
\includegraphics[width=8.0cm]{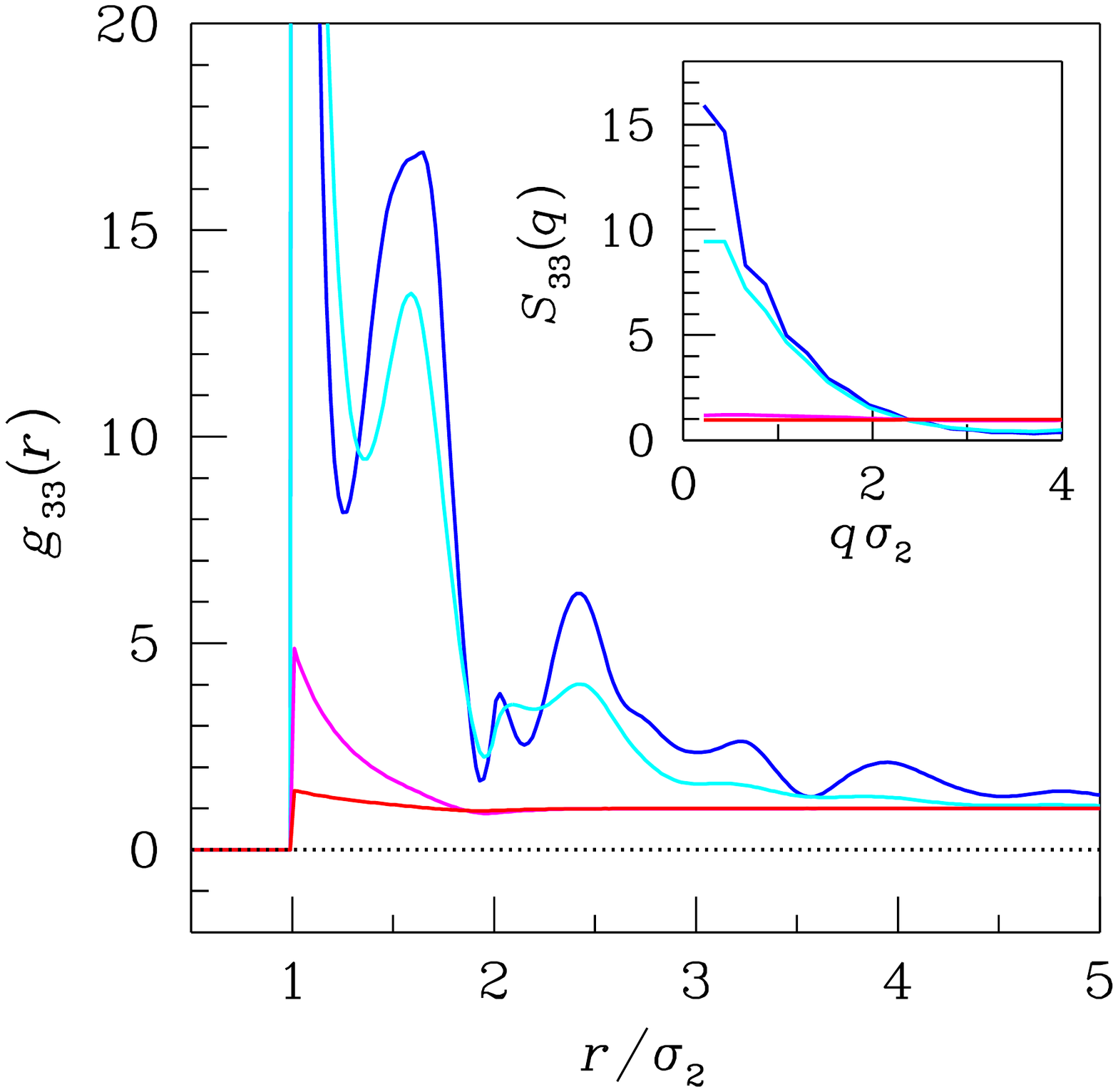} &
\includegraphics[width=8.0cm]{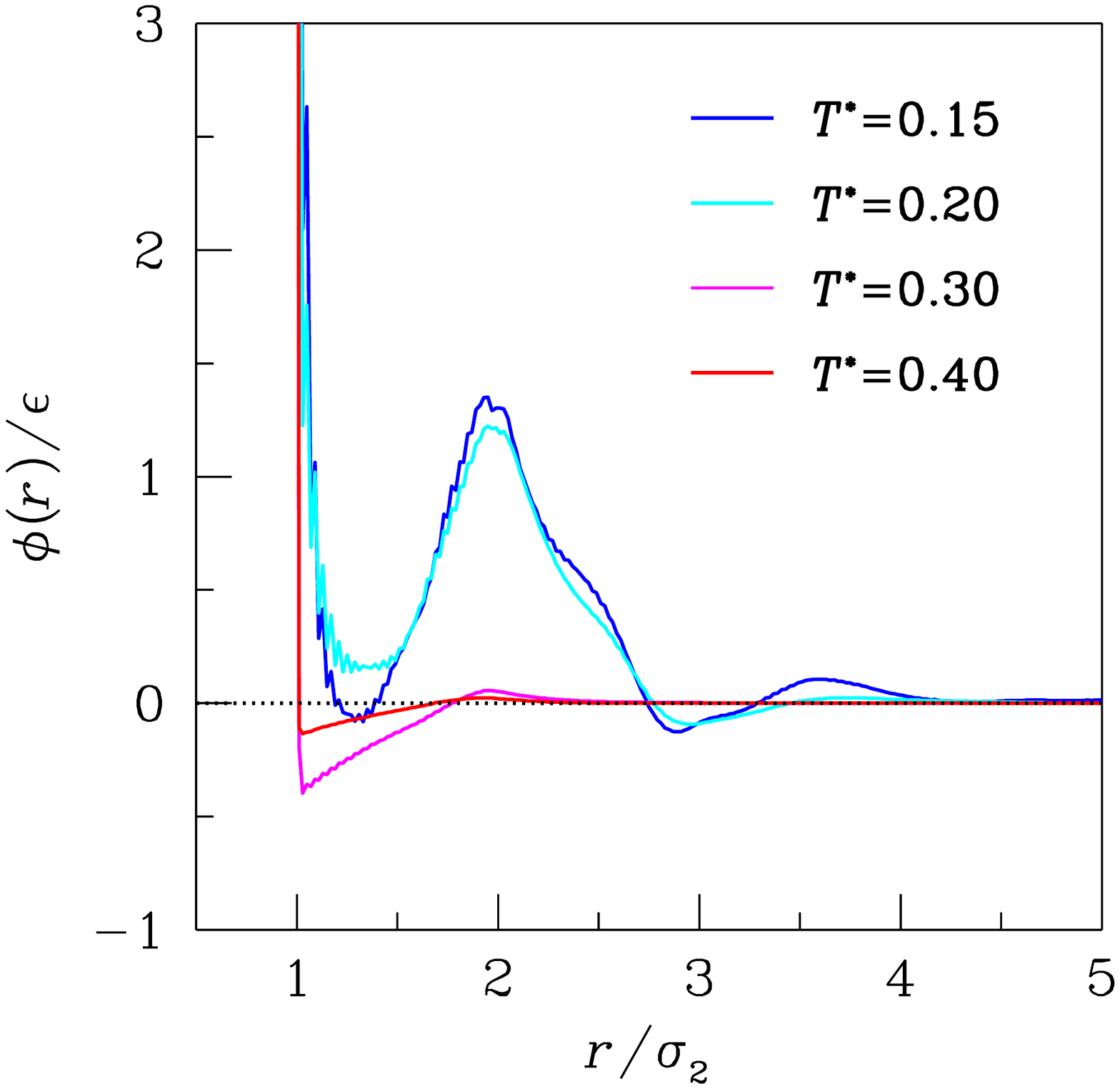}
\end{tabular}
\caption{Left: $g_{33}(r)$ (with corresponding $S_{33}(q)$ in the inset) for $\chi=33\%$ and various $T^*$ (see the legend). Right: guest-guest potential $\phi(r)$ obtained from the HNC inversion of $g_{33}(r)$.}
\label{fig8}
\end{center}
\end{figure*}

A more complete set of the relevant RDFs of the system is plotted in Fig.\,9 for $T^*=0.10$.  Clearly, the simulated system is far from being homogeneous at low temperature, which may cast doubts on the opportunity to use the RDF for investigating the model structure. However, at least the short-range structure ({\it i.e.}, the radial structure within distances smaller than the cross radius of a typical cluster) is well accounted for by the conventional RDF. As a first comment, we note that the development of an enormously large $g_{33}$ value at contact is the most distinct signature of the formation of sphere aggregates at low temperature. This is evident for all concentrations, even the higher ones. These aggregates are more extended for intermediate $\chi$ values, where the short-distance structure in $g_{33}$ is richer. As to the $g_{13}(r)$ function, its short-distance profile is sharper for the lowest concentrations, where the number of dimers that are in close contact with the same sphere is higher. Instead, the somewhat high third-neighbor peak at intermediate concentrations is another signature of the existence of more extended aggregates of dimers and guests for such $\chi$ values.

%
%
\begin{figure*}
\begin{center}
\begin{tabular}{cc}
\includegraphics[width=8.0cm]{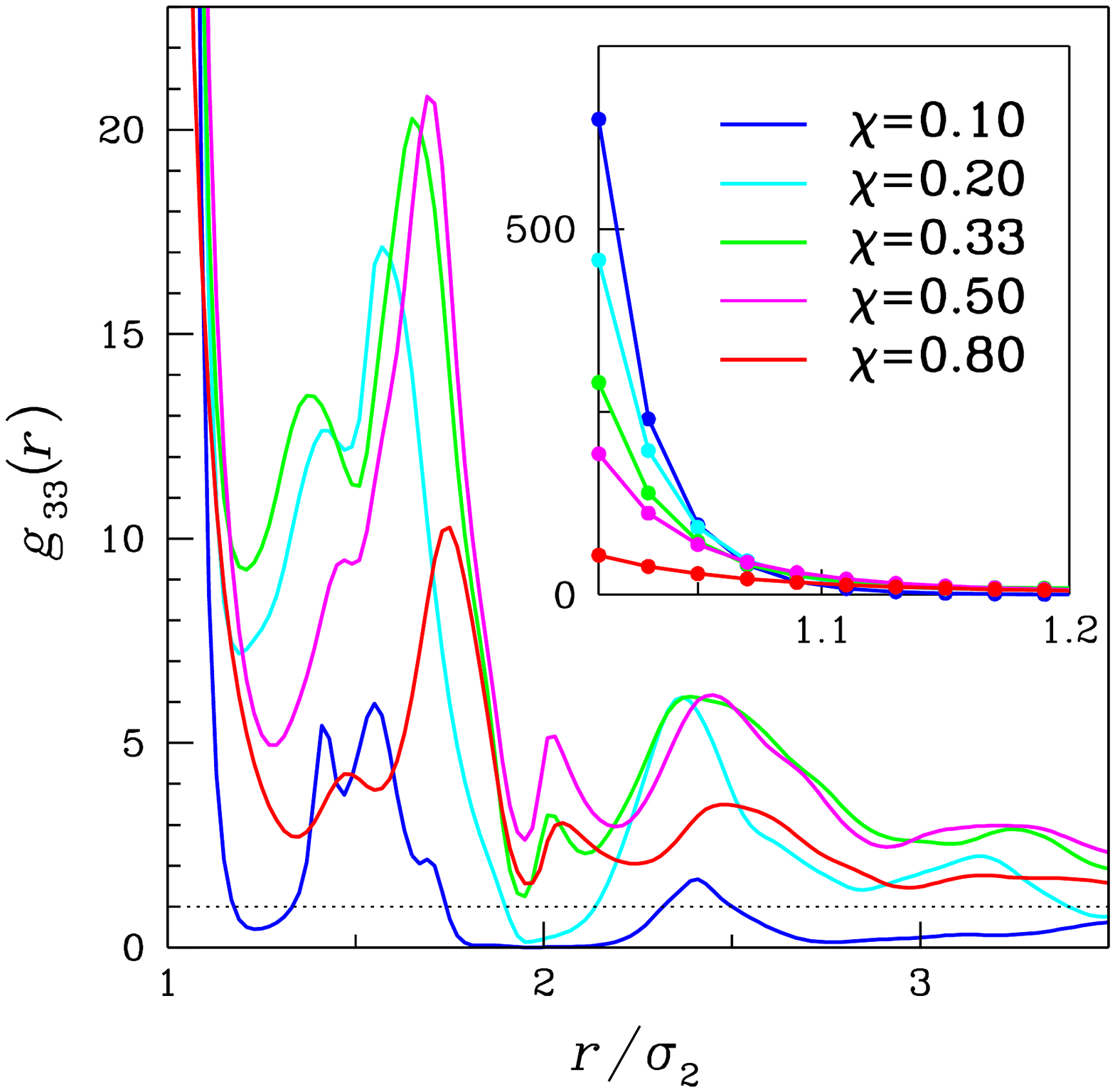} &
\includegraphics[width=8.0cm]{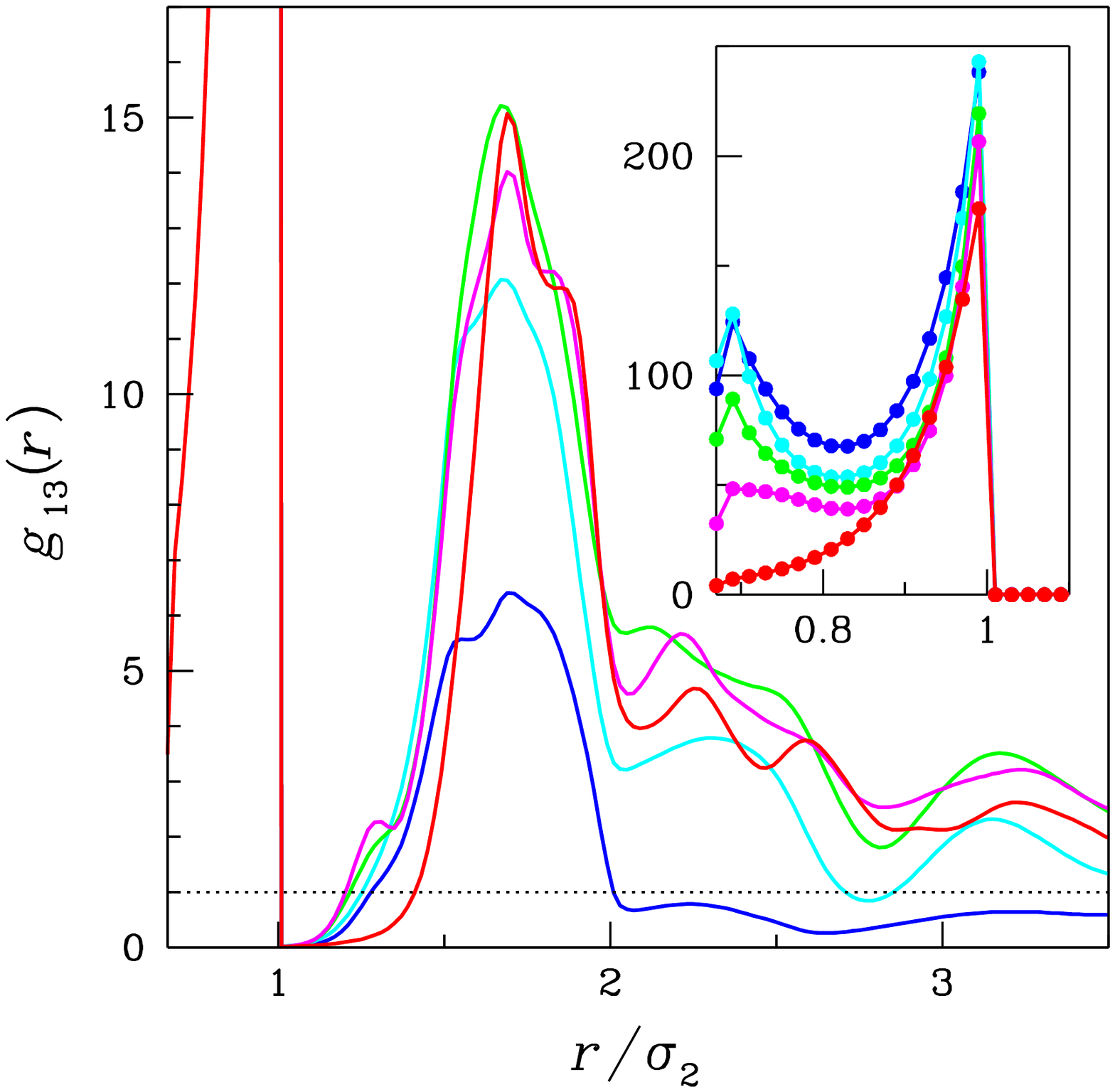}
\end{tabular}
\caption{$g_{33}(r)$ (left) and $g_{13}(r)$ (right) for $T^*=0.10$ and various $\chi$ (in the legend). Insets show a magnification of the short-distance region.}
\label{fig9}
\end{center}
\end{figure*}

Finally, we have computed the second virial coefficient $B_2$ of the mixture as a function of the temperature, seeking for its vanishing at fixed $\chi$. The value of $T$ where $B_2=0$ (the Boyle temperature $T_{\rm B}$) gives an indication of the threshold below which attractive forces start to be effective in the system. The expression for $B_2$ is:
\be
B_2=(1-\chi)^2B_2^{\rm DD}+2\chi(1-\chi)B_2^{\rm DS}+\chi^2B_2^{\rm SS}\,, 
\label{eq3-1}
\ee
where in the rhs the partial contributions to $B_2$ are from a dimer pair (DD), a dimer and a sphere (DS), and a sphere pair (SS; for hard spheres, $B_2^{\rm SS}=(2/3)\pi\sigma_3^3$). For instance, the dimer-dimer contribution reads:
\be
B_2^{\rm DD}=-\frac{1}{32\pi^2V}\int{\rm d}^3R_1{\rm d}^2\Omega_1{\rm d}^3R_2{\rm d}^2\Omega_2\left(e^{-\beta U_{12}}-1\right)\,, 
\label{eq3-2}
\ee
where, {\it e.g.}, $R_1$ and $\Omega_1$ are the coordinates defining the position and orientation of dimer 1, and $U_{12}$ denotes the interaction energy between dimers 1 and 2. We have computed $B_2^{DD}$ and $B_2^{DS}$ by Monte Carlo integration (see Ref.\,\cite{Yethiraj2} for details). In Fig.\,10 we report our results for $T_{\rm B}$ as a function of $\chi$ ($B_2$ has been computed in steps of $\Delta T^*=0.01$). We see that the Boyle temperature is systematically higher than the temperature where clusters first occur on cooling ($T^*\approx 0.25$). Moreover, the maximum of $T_{\rm B}(\chi)$ at $\chi\simeq 50\%$ will conform with the expectation that the larger the clusters, the wider is their range of stability.

The $B_2$ data in Fig.\,10 reveal that attractive forces between spheres and dimers become increasingly strong upon cooling. A large negative $B_2$ would suggest a tendency of the mixture to phase separate at low $T$; on the other hand, we may look at the sign of the second virial coefficient relative to spheres only, as computed from the effective interaction between spheres, $\phi(r)$ in Fig.\,8. From the right panel, which reports $\phi(r)$ for $\chi=33\%$ at various temperatures, we see that the effective potential is largely positive for $T^*=0.15$, implying that the corresponding $B_2$ is positive as well. Hence, we surmise that the cluster fluid is stable and phase separation, if any, would only occur for temperatures below 0.15.

%
%
\begin{figure}
\begin{center}
\includegraphics[width=8.0cm]{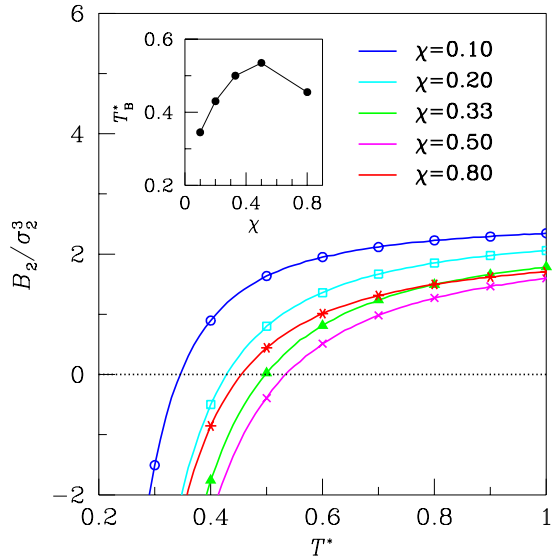}
\caption{$B_2$ vs. $T^*$ for various $\chi$ (see the legend). In the inset, $T_{\rm B}^*$ is plotted as a function of $\chi$.}
\label{fig10}
\end{center}
\end{figure}

\subsection{Evidence of lamellar structure at low temperature}

We found an unexpected behavior when relaxing the $\chi=50\%$ system for $T^*=0.15$. After about $5\times 10^8$ cycles, the energy dropped below the $T^*=0.10$ level (see Fig.\,11 top panel). Looking at the system snapshot beyond the crossing point, a new type of self-assembly emerges (Fig.\,12) where most of the particles are arranged into an ordered lamellar structure, that is a planar arrangement where the guest spheres are bound to a double layer of dimers. In Fig.\,12 the observed lamellae are actually three, fused together along an edge, and we also see a short straight tube floating in the simulation box. A clue to how particles are arranged in a lamella can be obtained from the distribution of the angle $\alpha$ between two nearest guest-guest bonds.

%
%
\begin{figure}
\begin{center}
\includegraphics[width=8.0cm]{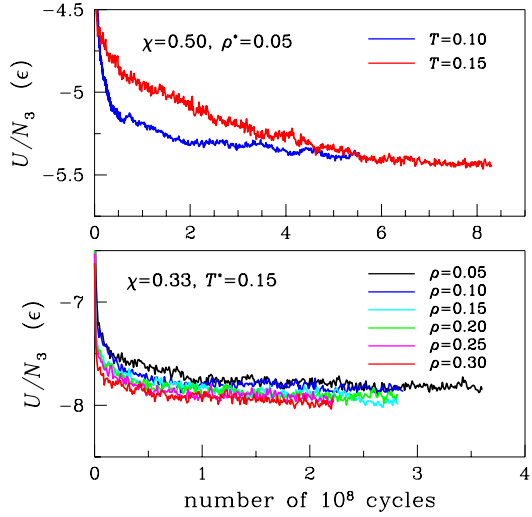}
\caption{MC evolution of the potential energy $U$ per sphere under different concentration, temperature, and density conditions.}
\label{fig11}
\end{center}
\end{figure}

%
%
\begin{figure}
\begin{center}
\includegraphics[width=10.0cm]{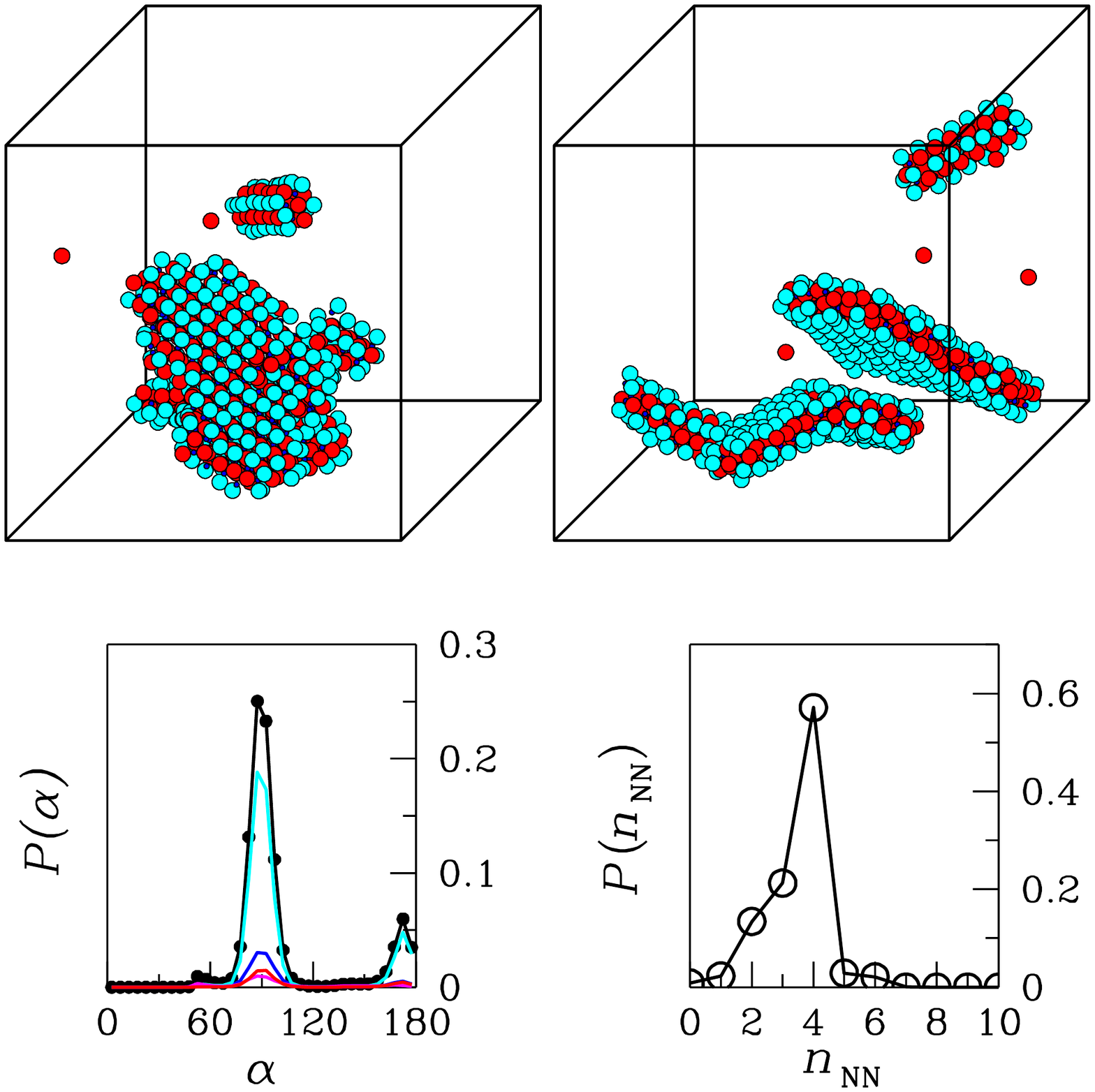}
\caption{Properties of the mixture for $\chi=50\%$ and $T^*=0.15$. The top panels report the same configuration as seen from different perspectives. The total number of guests is 400: 363 spheres belong to the lamellar aggregate, while 33 stay in the tube; finally, four spheres are isolated. Bottom: distribution of the angle $\alpha$ formed by two nearest guest-guest bonds (left, same notation as in Fig.\,6) and the distribution of the sphere coordination number $n_{\rm NN}$ (right).}
\label{fig12}
\end{center}
\end{figure}

Now, the leading maximum by far occurs for $\alpha=\pi/2$ (left-bottom panel of Fig.\,12), suggesting a square arrangement of guest spheres in the narrow space between the layers occupied by the dimers. A closer scrutiny of the configuration in Fig.\,12 reveals that spheres are actually distributed in two square sublattices that are slightly displaced one relative to the other (the nearest neighbors of a sphere are found in the other sublattice). Indeed, the $n_{NN}$ distribution reaches its maximum for a number of neighbors equal to four (Fig.\,12, right-bottom panel), which is consistent with the proposed picture. The structure of the tube is different: four rows of spheres are alternated to four rows of dimers, all being wrapped around another chain of spheres (an arrangement somehow reminiscent of that of metal atoms in nanowires~\cite{Tosatti2}). The significance of this outcome is that a lamellar solid coexisting with vapor competes for stability at low $T$ with a fluid of clusters, at least for guest concentrations close to $50\%$. The present scenario bears a strong similarity to what occurs in a one-component system of one-patch spheres~\cite{Preisler}, where the cluster fluid is metastable at low temperature.

To summarize, for $T^*=0.15$ the simulated sample eventually succeeded to reach equilibrium, thanks to a non-negligible particle diffusivity, and apparently separates into vapor and lamellar solid, as it does the system studied in Ref.\,\cite{Preisler} in a range of thermodynamic parameters (see Fig.\,10 therein). For the lower temperature of $T^*=0.10$, clusters are still formed very early in the system but they did not subsequently evolve, probably for lack of time.

Even though we are not in a position to draw an accurate phase diagram, we attempt a sketch in Fig.\,13 for $\rho^*=0.05$. Similarly as argued for $\chi=50\%$, for $T^*=0.10$ we conjecture the metastability of clusters throughout the whole range of guest concentrations (red symbols in Fig.\,13), especially considering that the average potential energy seems to still evolve in such conditions (see Fig.\,1).

The present results can be compared with those obtained in Ref.\,\cite{Slyk}, where a mixture of spheres and spherical patchy particles with two or more patches is studied in two dimensions by MC simulation. Leaving aside the question of the different dimensionality, the strongest similarity with our mixture is reached for patchy particles with two equal wide patches. In that case, at relatively high density various types of vapor-solid separation are reported at low temperature, similarly as found in our equimolar mixture.

%
%
\begin{figure}
\begin{center}
\includegraphics[width=8.0cm]{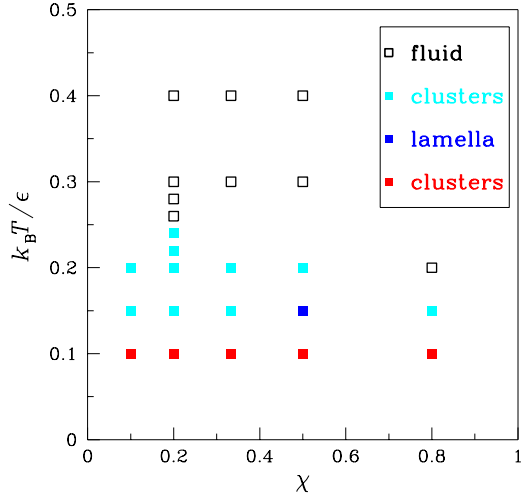}
\caption{Putative phase diagram of the mixture for $\rho=0.05$. Cyan and red symbols denote stable and possibly metastable clusters, respectively.}
\label{fig13}
\end{center}
\end{figure}

\subsection{Gelation}

A final question to address concerns the existence of a sol-gel transition (gelation) in a mixture of heteronuclear dimers and spherical guests. Gelation is a structural transition (the mean cluster size diverges within finite time)~\cite{Flory,deGennes,Lu}. Colloidal particles with a mutual attraction much stronger than the thermal energy exhibit a fluid-to-solid gelation transformation at a critical, temperature-dependent volume fraction~\cite{Zaccarelli}. The sol-gel transition is manifested in the onset of a {\em spanning cluster}, which gives rise to the divergence of viscosity as the transition point is approached from the sol phase, and to a vanishing elastic modulus as the transition is approached from the gel phase (see, {\it e.g.}, Ref.\,\cite{Stauffer}). A percolating, or spanning cluster has at least one particle on each of the six faces of the simulation box. Like percolation, gelation has many of the hallmarks of a critical phase transition. As the gelation point is approached, the sol viscosity, the shear modulus, and the mean cluster size all exhibit power-law behavior with characteristic exponents. Gelation is accompanied by kinetic arrest due to crowding of clusters; like the glass transition, gelation is akin to a jamming transition.

%
%
\begin{figure*}
\begin{center}
\begin{tabular}{cc}
\includegraphics[width=5.0cm]{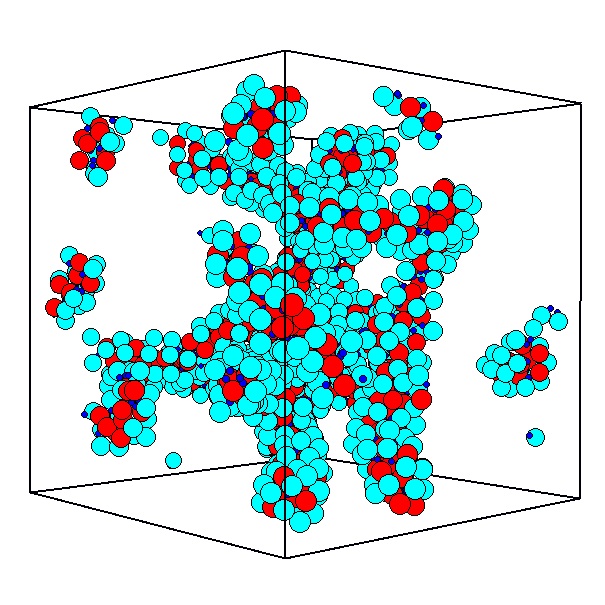} &
\includegraphics[width=5.0cm]{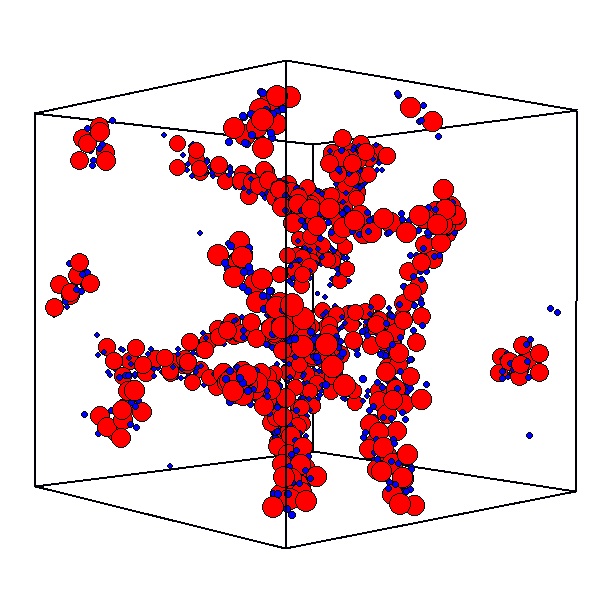}
\end{tabular}
\caption{Final system configuration for $\chi=33\%,T^*=0.15$, and $\rho^*=0.15$. A percolating structure formed by all particles in the system is clearly visible on the left. On the right, the large monomers have been removed in order to highlight the cluster backbone. Note that the disconnected parts are an artifact of periodic boundary conditions.}
\label{fig14}
\end{center}
\end{figure*}

According to the evidence presented so far, no gelation occurs in our system for $\rho^*=0.05$, {\it i.e.}, no sign is ever found of a percolating network of guests. But it might as well be that the simulated system is simply too dilute (and $T^*=0.10$ is too low a temperature) for gelation to occur within affordable time. To check this intuition we have gradually increased the density of the $\chi=33\%$ system, up to $\rho^*=0.30$, while keeping the temperature fixed at $T^*=0.15$. As expected, the asymptotic value of the potential energy slightly decreases upon compression (see Fig.\,11 bottom panel). As $\rho$ increases, the clusters occurring in simulation become larger and larger until, for $\rho^*\approx 0.15$, a single cluster encompassing almost all system particles appears for the first time and maintains practically unaltered in the subsequent part of the MC evolution (see Fig.\,14 for a visual representation of the gelified system; in the case shown, only one sphere out of a total of 400 is not part of the spanning cluster). It goes without saying that we do not have the necessary resolution in density and temperature to see whether the characteristics of gelation for the present model do actually conform with the generic expectation of a ``critical'' transition. Finally, both $n_{NN}$ and bond-angle distributions of the gelified system are found to be relatively independent of the particle density (be it $\rho^*=0.15$ or larger), being quite similar to the corresponding distributions for $\rho^*=0.05$.

A final issue concerns the question of whether the observed network is a genuine gel or rather a percolating arrested state on the way to phase separation between vapor and (tubular) solid. It is commonly believed (see e.g. Refs.\,\cite{Zaccarelli,Foffi}) that the difference between the two scenarios lies in the relative position of the glass-transition line and the binodal line in the $\rho$-$T$ diagram. Should the glass line meet the phase-coexistence locus on its high-$\rho$ branch, the gel formed would be an arrested two-phase coexistence state. However, since we have neither a precise idea of the overall mixture phase diagram, nor any knowledge of the location of the glass-transition line, the question about the nature of the percolating cluster remains presently unsettled.

\section{Conclusions}

We have investigated the self-assembly of a mixture of heteronuclear dimers and spherical particles. All interactions are of hard-sphere type; in addition, a square-well attraction is present between a guest sphere and the smaller particle in a dimer. We have fixed the sphere diameter to that of the larger monomer, with the purpose to describe a colloidal mixture where all solute particles are similar in size. Using Monte Carlo simulation, we have mapped the full emergent behavior of the system as a function of the sphere concentration $\chi$, even far beyond the low-$\chi$ encapsulation regime.

Our findings can be summarized as follows. At low temperature, the tendency of particles to reduce internal energy produces quite distinctive organized structures in an overall dilute mixture. These structures have the character of spheroidal clusters for low or very high concentrations, while looking more like curved tubes for intermediate $\chi$ values. These aggregates are rather different from the precipitates that form when small molecules attract each other in a poor solvent, as well as from the liquid clusters spontaneously arising in a supercooled vapor. The difference mainly originates from the fact that in an initially homogeneous mixture of dimers and spheres the formation of a large number of bound pairs soon results in the saturation of the attraction, just for steric reasons. Indeed, as the simulation goes on an increasing number of binding sites get entrapped in the interior of aggregates, this way becoming unavailable for other bonds. As a result, the initial stage of fast cluster growth comes to a stop and clusters become relatively stable (at low temperature, the binding energy is so strong that cluster particles only hardly get unstick). This is reflected in the shape of the effective guest-guest potential, featuring a short-range attraction and a long-range repulsion. At intermediate concentrations, the number of spheres and binding sites exposed to the surface of the clusters are both sufficiently high during the growth stage that big clusters eventually arise. However, only when the overall system density is high enough (at least 0.15 in reduced units) a spanning structure (gel) is formed in relatively short time.

In one case only, that is $\chi=50\%$ and a temperature not so low that particle mobility is significantly suppressed, Monte Carlo relaxation to equilibrium had an unexpected outcome: the formation of an ordered lamellar structure with a square arrangement of guest spheres and dimers. The very same existence of this structure indicates a low-temperature competition between clusters and crystal-vapor separation. It is reasonable to expect that, while clusters are kinetically favored at low $T$, the separation between lamellar crystal and vapor is the solution adopted by the system in stable equilibrium.

In a forthcoming paper, we plan to analyze in detail the case of guest spheres much larger than dimers, focusing our attention on the competition between clusterization and (macro-)separation into a guest-rich and a guest-poor phase~\cite{Munao:capsule}. An extension of the dimeric hard-sphere model to include non-additive hard sphere effects~\cite{Pellicane2} is also in program.

\end{document}